# Performance Investigation of Feature Selection Methods and Sentiment Lexicons for Sentiment Analysis

Anuj Sharma
Information Systems Area
Indian Institute of Management,
Indore, India

Shubhamoy Dey
Information Systems Area
Indian Institute of Management,
Indore, India

## ABSTRACT

Sentiment analysis or opinion mining has become an open research domain after proliferation of Internet and Web 2.0 social media. People express their attitudes and opinions on social media including blogs, discussion forums, tweets, etc. and, sentiment analysis concerns about detecting and extracting sentiment or opinion from online text. Sentiment based text classification is different from topical text classification since it involves discrimination based on expressed opinion on a topic. Feature selection is significant for sentiment analysis as the opinionated text may have high dimensions, which can adversely affect the performance of sentiment analysis classifier. This paper explores applicability of feature selection methods for sentiment analysis and investigates their performance for classification in term of recall, precision and accuracy. Five feature selection methods (Document Frequency, Information Gain, Gain Ratio, Chi Squared, and Relief-F) and three popular sentiment feature lexicons (HM, GI and Opinion Lexicon) are investigated on movie reviews corpus with a size of 2000 documents. The experimental results show that Information Gain gave consistent results and Gain Ratio performs overall best for sentimental feature selection while sentiment lexicons gave poor performance. Furthermore, we found that performance of the classifier depends on appropriate number of representative feature selected from text.

## General Terms
Sentiment Analysis, Machine Learning Techniques, Classification.

## Keywords
Sentiment Analysis, Feature Selection, Sentiment Lexicon, Classification

## 1. INTRODUCTION
The popularity and reach of Internet and, specifically the Web 2.0 social media have changed the way we generate, share and utilize the information. The audience (i.e., the receivers of the information) has gone far away from just consuming the available content to actively annotating, commenting and contributing to the content. Web 2.0 has provided new mediums where people can express their interest, attitude, views, opinion and feedback. Social networking sites, blogs, discussion forums, microblogs and content sharing sites provides such facilities to disseminate user generated information [1]. The information on these mediums is freely available online in a text format and contains voice of the public. For example, many Internet based discussion forums and review sites enable people to express their views about a product or service. While analyzing such consumer-generated online text content offers tremendous business opportunities in term of finding consumer preferences and getting their feedback [2]. This presents an opportunity for business organizations to better understand and efficiently respond to the consumers by processing their unsolicited feedback.

Due to the huge and ever-growing information on the Web, which is highly unstructured and scattered, we are now barely able to access and utilize the information. Multiple solutions have been proposed to solve this problem, and they are mainly specialized in factual information retrieval (IR), data mining (DM) or more specifically text mining (TM), natural language processing (NLP) and machine learning techniques (ML) [3]. Sentiment analysis (often referred as opinion mining) is a recent area of research where we apply these advanced techniques to process vast amounts of user generated text content. The purpose of sentiment analysis is to determine the overall opinion or attitude, in term of positive or negative, expressed in text available over Internet. Sentiment analysis may be as simple as the basic sentiment based categorization of text documents to more complex and advanced procedures to extract opinion at different granularity levels [4]. Sentiment classification differs from text categorization in term of criterion of classification, which is the opinion or view expressed, instead of topic or frequent features in the text [5, 6].

Sentiment based classification of text documents is more challenging tasks than topic based classification since this involves discrimination based on opinions, feelings, and attitudes contained in text. The opinionated or subjective text on the Web is often non-structured or semi-structured. Furthermore, due to high dimensionality of the text content feature selection is a crucial problem [7, 8]. The sentiment features are not expressed objectively and explicitly, and usually are hidden in a large pool of subjective text. Therefore, the text sentiment classification requires deeper analysis and understanding of textual features.

Applying effective and efficient feature selection can enhance the performance of sentiment analysis in term of accuracy and time to train the classifier. This paper explores applicability of feature selection methods for sentiment based classification and investigates their performance in term of recall, precision and accuracy. In this paper, five feature selection methods (Document Frequency, Information Gain, Gain Ratio, Chi Squared, Relief-F) and three popular sentiment feature lexicons (HM, GI and Opinion Lexicon) are investigated on movie reviews corpus with a size of 2000 documents. The movie review dataset comprising reviews of movies from Internet Movie Database (IMDb). Support vector machine (SVM) classifier is used for sentiment classification experiments [9]. SVM have shown good track record in text classification [5]. As a whole this paper focuses on two different sentiment analysis methods. The first is purely machine-learning based method that is to classify textual movies reviews into either a positive or negative class. The second method exploits natural language processing and uses different opinion dictionary (sentiment lexicons) and to detect words carrying opinion in the corpus and then to predict opinion expressed in the text.

The rest of this paper is organized as follows: Section 2 presents related work on sentiment analysis. Feature selection and sentiment lexicon based method are described in Section 3. Experimental results are given in Section 4. Finally Section 5 concludes this paper.

## 2. RELATED WORK

Earlier studies in this domain focused on assigning sentiments to documents [5, 6] utilizing machine learning techniques which usually attempted training a sentiment classifier based on frequent terms in the documents. The others studies focused on more specific tasks that was finding the sentiments or semantic orientation of words [10, 11]. Since the early days of sentiment analysis, machine learning has been the most exploited technique to tackle the relevant problems. The pioneer work on sentiment classification by Pang and Lee (2002) utilized naive bayes (NB), maximum entropy (ME) and support vector machines (SVM). As per their results, SVM showed the best performance, while NB was the least precise out of the three. On the movies reviews collection from Internet Movie Database (IMDb), 82.9 % accuracy was achieved [5]. The work by Dave et al. (2003) used Laplacian smoothing for NB, which enhanced its accuracy to 87% for a product reviews dataset [12]. To improve the precision of NB, Pang and Lee (2004) performed subjectivity identification as a pre-processing step to sentiment analysis [13].

To address the problem of multi-class sentiment analysis, Pang and Lee (2005) proposed to use SVM in multi-class one-versus-all (OVA) and regression (SVR) modes, combining them with metric labeling. The work demonstrated that a combination of SVM with other unsupervised classification methods resulted in better precision [14]. Modeling relationships and agreement between authors in political texts were also utilized as an extension to the SVM approach [15].

The performance of machine learning based sentiment analysis depends upon quality and quantity of training corpus. The gold standard training data is always much less than the amount of unlabeled data. A semi-supervised learning technique was also proposed which operated on a graph of both labeled and unlabeled corpus and this approach exhibited better performance than SVR [16]. Chen et al. (2006) used decision trees, SVM and NB for sentiment based classification of reviews of book, *The Da Vinci Code*, and achieved 84.59% accuracy [17]. The study showed opinions on one topic with different graph structures. Chen et al. (2006) presented term clusters with polarity information, words coordination, and decision tree based review representation.

Boiy et al. (2007) performed experiments using SVM, naive Bayes multinomial and maximum entropy on movie and car brands review. The best accuracy of the study was up to 90.25% [18]. The similar movie reviews dataset was used for experiment by Annett and Kondrak (2008). Different approaches like SVM, NB, alternating decision tree and lexical (WordNet) based approach were utilized for sentiment analysis and greater than 75% accuracy was achieved [19]. Tan and Zhang (2008) compared different feature selection (Mutual Information, Information Gain, Chi Squared and Document Frequency) and learning methods (centroid classifier, K-nearest neighbor, window classifier, Naïve Bayes and SVM) in extracting opinion from Chinese documents. Information Gain performed the best for sentimental terms extraction and SVM exhibited the best performance for sentiment classification [20].

In a study by Dasgupta and Ng (2009), a weakly-supervised sentiment classification algorithm was proposed. User feedbacks were provided on the spectral clustering process in an interactive mode to ensure that text is clustered along the sentiment dimension [21]. A study on travel blogs was done by Ye et al. (2009) based on NB, SVM and the character based N-gram model. SVM performed best in the study and gave 85.14% best accuracy [22]. Support Vector Machine was also exploited for sentiment analysis by Paltoglou & Thelwall (2010) for proposing a combined approach which gave up to 96.90% accuracy on movie reviews dataset [23]. Xue Bai (2011) experimented with heuristic search-enhanced Markov blanket model and applied SVM for mining consumer sentiments from online text [24]. Kang et al. (2011) improved Naïve Bayes to use on restaurant reviews and obtained 83.6% accuracy on more than 6000 documents [25].

Sentiment lexicon or semantic orientation based approaches included some early work like Hu and Liu (2004) which used the adjective synonym and antonym sets from WordNet to evaluate the semantic orientation of adjectives [26]. The adjective set is used as sentiment dictionary or sentiment lexicon for semantic orientation based sentiment analysis. These work usually involved the manual or semi-manual construction of semantic orientation word lexicons developed by word sentiment classification techniques [10]. Some studies confirmed that restricting features to adjectives for word sentiment classification would improve performance [27]. Moreover, other showed that most of the adjectives, adverb, and a small set of nouns and verbs can acquire semantic orientation [28]. While the machine learning based approaches provided better classification accuracy, but required a lot of training time and pre-classified training corpus. Moreover Semantic orientation based approached did not gave good performance, but they returned results quickly.

Extracting the best feature in appropriate number is a crucial task in machine learning-based sentiment classification. Different studies have attempted to extract complex features and have proposed comparative feature selection methods [7]. Most of the existing research literature focus on simple features, including single words [13], character N-grams [5], and word N-grams [13, 14] like bigrams and trigrams [12], or the combination of aforementioned features. Some other studies has adopted different feature selection methods like log likelihood tests [17], Fisher's discriminant ratio [7], information gain and CHI statistics [20]. This paper investigates performance of different feature selection methods from data mining research for the purpose of sentiment based classification.

## 3. METHODOLOGY

This section explains the methodology of the proposed study as presented in Figure 1. First, the review documents were collected and pre-processed with basic natural language processing techniques like word tokenization, stop word removal and stemming. The residual tokens were arranged as per their frequencies or occurrences in whole documents set. Then different feature selection methods were utilized to pick out top n-ranked discriminating attributes for training the classifier and sentiment based classification. The sentiment lexicons were used to select the sentiment expressing features from the review texts. Support Vector Machine (SVM) is applied to evaluate the effectiveness of five feature selection methods on the performance of sentiment classification.

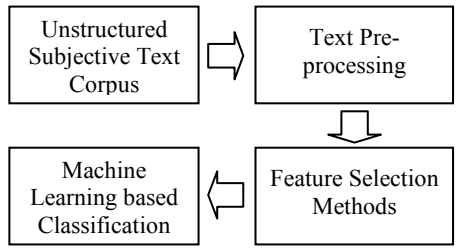

**Fig 1: The Methodology of the Proposed Study**

## 3.1 Feature Selection Methods

Feature selection methods reduce the original feature set by removing irrelevant features for text sentiment classification to improve classification accuracy and decrease the running time of learning algorithms. We have investigated performance of five commonly used feature selection methods in data mining research, i.e., DF, IG, CHI, GR and Relief-F. All these feature selection methods compute a score for each individual feature and then select top ranked features as per that score.

### 3.1.1 Document Frequency (DF)
Document Frequency measures the number of documents in which the feature appears in a dataset. This method removes those features whose document frequency is less than or greater than a predefined threshold frequency. Selecting frequent features will improve the likelihood that the features will also be comprised by prospective future test cases. The basic assumption is that both rare and common features are either non-informative for sentiment category prediction, or not impactful to improve classification accuracy [29]. Research literature shows that this method is simplest, scalable and effective for text classification [20].

### 3.1.2 Information Gain (IG)
Information gain is utilized as a feature (term) goodness criterion in machine learning based classification [7, 20, 29]. It measures information obtained (in bits) for class prediction of an arbitrary text document by evaluating the presence or absence of a feature in that text document. Information Gain is calculated by the feature's contribution on decreasing overall entropy. The expected information needed to classify an instance (tuple) for partition $D$ or identify the class label of an instance in $D$ is known as entropy and is given by:

$$Info(D) = -\sum_{i=1}^{m}(P_i)\log_2(P_i) \quad (1)$$

Where $m$ represents the number of classes (m=2 for binary classification) and $P_i$ denotes probability that a random instance in partition $D$ belongs to class $C_i$ estimated as $|C_i, D| / |D|$ (i.e. proportion of instances of each class or category). A log function to the base 2 justifies the fact that we encode information in bits.

If we have to partition (classify) the instance in $D$ on some feature attribute $A$ $\{a_1,..., a_v\}$, $D$ will split into $v$ partitions set $\{D_1, D_2,..., D_v\}$.

The amount of information in bits, we still require for an exact classification is measured by:

$$Info_A(D) = -\sum_{j=1}^{v}\frac{|D_j|}{|D|} \times Info(D_j) \quad (2)$$

Where $|D_j|/|D|$ is the weight of the $j^{th}$ partition and $Info(D_j)$ is the entropy of partition $D_j$.

Finally Information gain by partitioning on $A$ is

$$Information\ Gain(A) = Info(D) - Info_A(D) \quad (3)$$

We select the features ranked as per the highest information gain score. We can optimize the information needed or decrease the overall entropy by classifying the instances using those ranked features.

### 3.1.3 Gain Ratio (GR)
Gain Ratio enhances Information Gain as it offers a normalized score of a feature's contribution to an optimal information gain based classification decision. Gain Ratio is utilized as an iterative process where we select smaller sets of features in incremental fashion. These iterations terminate when there is only predefined number of features remaining. Gain ratio is used as one of disparity measures and the high gain ratio for selected feature implies that the feature will be useful for classification. Gain Ratio was firstly used in decision tree (C4.5), and applies normalization to information gain score by utilizing a split information value [30].

The split information value corresponds to the potential information obtained by partitioning the training data set $D$ into $v$ partitions, resulting to $v$ outcomes on attribute $A$:

$$SplitInfo_A(D) = -\sum_{j=1}^{v}\frac{|D_j|}{|D|} \times \log_2\frac{|D_j|}{|D|} \quad (4)$$

Where high *SplitInfo* means partitions have equal size (uniform) and low *SplitInfo* means few partitions contains most of the tuples (peaks). Finally the gain ratio is defined as:

$$Gain\ Ratio(A) = Information\ Gain(A) / SplitInfo(A) \quad (5)$$

### 3.1.4 CHI statistic (CHI)
The Chi Squared statistic (CHI) measures the association between the word feature and its associated class or category. CHI as a common statistical test represents divergence from the distribution expected (i.e. resultant partition) based on the assumption that the feature occurrence is perfectly independent of the class value [20, 29]. It is defined as,

$$CHI(t, c_i) = \frac{N \times (AD - BE)^2}{(A+E) \times (B+D) \times (A+B) \times (E+D)} \quad (6)$$

$$CHI_{\max}(t) = \max_i(CHI(t, c_i)) \quad (7)$$

Where $A$ is the frequency when $t$ and $C_i$ co-occur; $B$ represents counts when $t$ occurs without $C_i$. $E$ is the number representing events when $C_i$ occurs without $t$; $D$ is the frequency when neither $C_i$ nor $t$ occurs; $N$ represents total documents in the corpus. The CHI statistic will be zero if $t$ and $C_i$ are independent.

### 3.1.5 Relief-F Algorithm
The basic principle of Relief-F is to select feature instances at random, compute their nearest neighbors, and optimize a feature weighting vector to award more importance (weight) to features that discriminate the instance from neighbors of different classes [31]. Specifically, Relief-F attempt to evaluate a good estimation of weight $W_f$ from the following probabilities for weighting and ranking feature $f$:

$$W_f = P(different\ value\ of\ f\ |\ nearest\ instances\ from\ different\ class) - P(different\ value\ of\ f\ |\ nearest\ instances\ from\ same\ class) \quad (8)$$

## 3.2 Machine learning Classifier: Support Vector Machines
Support vector machines (SVMs) have been studied as highly effective text categorization technique that generally outperforms Naive Bayes [5, 13]. SVM attempts to seek a hyper-plane represented by vector that separates the positive and negative training vectors of documents with maximum margin. The problem of findings this hyperplane can be translated into a constrained optimization problem.

SVM utilizes the structural risk minimization principle that is based on the computational learning theory. SVM seeks a decision surface to separate the training data points into two classes (for binary classification) and makes classification decisions based on the support vectors that are selected as the only effective elements in the training set. The optimization of SVM (dual form) is to minimize:

$$\vec{\alpha}^* = \arg\min\left\{-\sum_{i=1}^{n}\alpha_i + \sum_{i=1}^{n}\sum_{j=1}^{n}\alpha_i\alpha_j y_i y_j \langle \vec{x}_i, \vec{x}_j \rangle\right\} \quad (9)$$

Subject to: $\sum_{i=1}^{n}\alpha_i y_i = 0;\ 0 \leq \alpha_i \leq C$ (10)

## 4. EXPERIMENT RESULTS
### 4.1 Datasets and Performance Evaluations
This study uses a data set of classified movie reviews prepared by Pang and Lee [13, 14]. The data set contains 1,000 positive and 1,000 negative reviews collected from Internet Movie Database (IMDb) and known as polarity dataset v2.0 or Cornell Movie Review Dataset. The sentiment lexicons used in this study are HM, GI and Opinion Lexicon. The HM dataset proposed by Hatzivassiloglou & McKeown (1997) is a popular sentiment lexicon which consists of 1336 adjectives, 657 positive and 679 negative [10]. The other sentiment lexicon is the GI dataset, which is a list of labelled words extracted from the General Inquirer lexicon. GI dataset includes 3596 adjectives, adverbs, nouns, and verbs, in which 1614 are positive and 1982 are negative. The Opinion Lexicon is adopted from [26].

To evaluate the performance of sentiment classification, this paper has adopted precision, recall and F-Measure as a performance measure.

$$\text{Recall} = \frac{\text{number of correct positive predictions}}{\text{number of positive examples}} \quad (11)$$

$$\text{Precision} = \frac{\text{number of correct positive predictions}}{\text{number of positive predictions}} \quad (12)$$

$$F - \text{Measure} = \frac{2 \times \text{Recall} \times \text{Precision}}{\text{Recall} + \text{Precision}} \quad (13)$$

### 4.2 Experimental Design
This paper has used Vector space model (feature vector model) for representing text documents of online reviews. Documents are represented as vectors and each dimension corresponds to a separate feature. Different feature selection methods were utilized to pick out top n-ranked discriminating attributes for training the classifiers where n was kept very small to very large (100 – 10000). Java based implementations on Microsoft Windows platform were used to implement all the feature selection methods. For experiments involving SVM, this paper employed radial basis function kernel as it gave better performance when tested empirically. All experiments were validated using 10-fold cross validation in which, the whole dataset is broken into 10 equal sized sets and classifier is trained on 9 datasets and tested on remaining dataset. This process is repeated 10 times and we take a mean accuracy of all fold.

### 4.3 Comparison and Analysis
Figure 2-4 shows performance of five feature selection methods in term of recall, precision and F-measure of the support vector machine classifier when trained and tested on the extracted features.

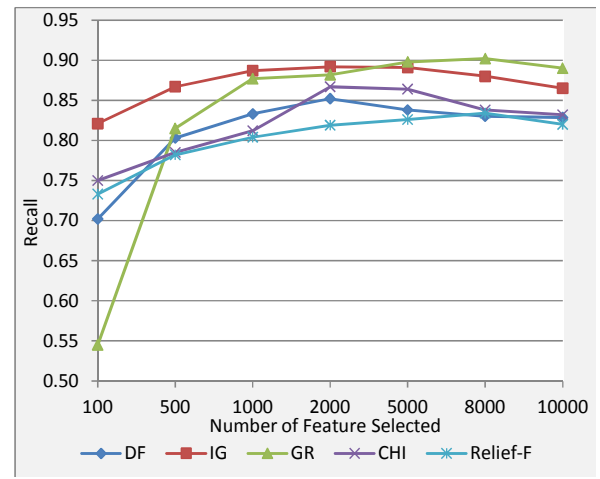

**Fig 2: Performance Comparison on Recall**

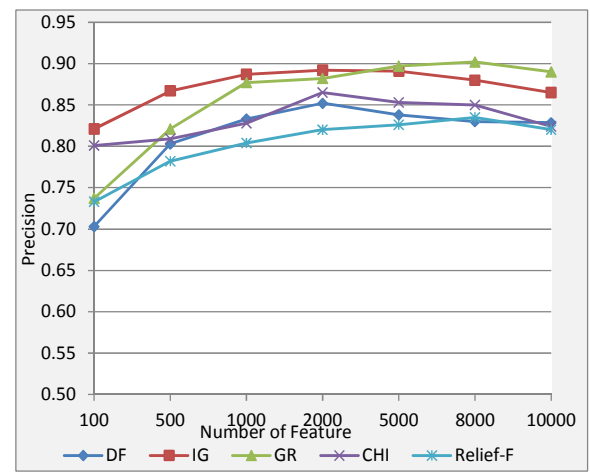

**Fig 3: Performance Comparison on Precision**

While comparing the recall and precision of classification with respect to feature selection methods, Gain Ratio (GR) gave the best results when the number of features selected were quite large (more than 5000). It gave best F-Measure as 0.901 with top 5000 features when used with SVM classifier. This indicated that GR is the best choice for feature selection method for studies related with sentiment analysis when we have to use large number of features. Research literature also supports this finding as, due to using a normalized score of a feature's contribution to a sentiment based classification decision, GR outperforms Information Gain feature selection [11]. The basic drawback found using GR is its sensitivity to number of feature selected as depicted from Figure 2 and 3.

When used with very less number of features, GR gave poor results as comparing to IG, CHI and Relief-F. Information Gain gave overall stable result and did not show any sensitivity to number of feature selected. Relief-F and DF gave poor results as compared to GR, IG and CHI.

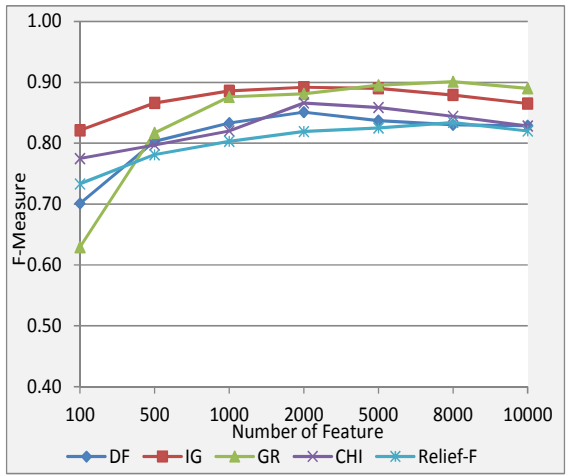

**Fig 4: Performance Comparison on F-Measure**

We have also trained and tested the SVM classifier on extracted features using HM, GI and Opinion Lexicon and the results are shown in the Table 1. We can see that the best F-Measure achieved by GI lexicon is 0.72 which is quite less than the performance of Gain Ratio or Information Gain based feature selection. The HM lexicon dataset performed even worse than GI and Opinion Lexicon. The domain specific nature of sentiment lexicon plays a vital part in affecting the performance of the classifier in sentiment based classification.

**Table 1. Comparison of Sentiment Lexicon with SVM**

| S. No. | Sentiment Lexicon | Recall | Precision | F-Measure |
|---|---|---|---|---|
| 1. | HM Lexicon | 0.49 | 0.64 | 0.55 |
| 2. | GI Lexicon | 0.739 | 0.76 | 0.75 |
| 3. | Opinion Lexicon | 0.675 | 0.723 | 0.70 |

## 5. CONCLUSION REMARKS

This paper explored applicability of feature selection methods for sentiment analysis and investigates their performance on classification in term of recall, precision and accuracy. Five feature selection methods (Document Frequency, Information Gain, Gain Ratio, Chi Squared, Relief-F) and three popular sentiment feature lexicons (HM, GI and Opinion Lexicon) are investigated on movie reviews corpus with a size of 2000 documents. The experimental results show that information gain gives stable performance for different number of features. In this study, Gain Ratio gave the best results for large number of sentimental features selection (more than 5000 features), while sentiment lexicons gave poor performance. Furthermore, we found that performance of the classifier depends on appropriate number of representative feature selected from text. We have shown that the appropriate number of features can be from 2000 to 8000 for better results. A promising direction for future work is to investigate the performance of feature selection methods on different machine learning classifiers and to evaluate the model for cross domain sentiment analysis with other domain than movie reviews.